\documentclass[reprint,amsfonts,showpacs,pas,prl,aps,english,superscriptaddress,float]{revtex4-1}
\usepackage{graphicx}
\usepackage{overpic}
\usepackage{color}
\usepackage{rotating}
\usepackage{txfonts}
\draft % marks overfull lines with a black rule on the right

\begin{document}

\title{Self-truncated ionization injection and consequent monoenergetic electron bunches in laser wakefield acceleration} %Title of paper

\author{Ming Zeng}
\affiliation{Key Laboratory for Laser Plasmas (Ministry of Education), Department of Physics and Astronomy, Shanghai Jiao Tong University, Shanghai 200240, China}
\author{Min Chen}
\email{minchen@sjtu.edu.cn}
\affiliation{Key Laboratory for Laser Plasmas (Ministry of Education), Department of Physics and Astronomy, Shanghai Jiao Tong University, Shanghai 200240, China}
\affiliation{Department of Mathematics, Institute of Natural Sciences, and MOE-LSC, Shanghai Jiao Tong University, Shanghai 20040, China}
\author{Zheng-Ming Sheng}
\email{zmsheng@sjtu.edu.cn}
\affiliation{Key Laboratory for Laser Plasmas (Ministry of Education), Department of Physics and Astronomy, Shanghai Jiao Tong University, Shanghai 200240, China}
\affiliation{Department of Mathematics, Institute of Natural Sciences, and MOE-LSC, Shanghai Jiao Tong University, Shanghai 20040, China}
\author{Warren B. Mori}
\affiliation{Department of Electrical Engineering, UCLA, Los Angeles, California 90095, USA}
\affiliation{Department of Physics and Astronomy, UCLA, Los Angeles, California 90095, USA}
\author{Jie Zhang}
\affiliation{Key Laboratory for Laser Plasmas (Ministry of Education), Department of Physics and Astronomy, Shanghai Jiao Tong University, Shanghai 200240, China}
\affiliation{Beijing National Laboratory of Condensed Matter Physics, Institute of Physics, CAS, Beijing 100190, China}
\date{\today}

\begin{abstract}
The ionization-induced injection in laser wakefield acceleration
has been recently demonstrated to be a promising injection scheme.
However, the energy spread controlling in this mechanism remains a
challenge because continuous injection in a mixed gas target is
usually inevitable.  Here we propose that by use of certain
initially unmatched laser pulses, the electron injection can be
constrained to the very front region of the mixed gas target,
typically in a length of a few hundreds micro meters determined by
laser-driven bubble deformation. Under some optimized conditions,
the injection region is well limited within $200\rm \mu m$ and the
electron beam with central energy of $383~\rm MeV$, energy spread
of $\Delta E_{\rm FWHM}/E = 3.33\%$, normalized emittance of
$3.12~\rm mm\cdot mrad$ and charge of $14.58~\rm pC$ can be
obtained according to particle-in-cell simulations. Both
multi-dimensional simulations and theoretical analysis illustrate
the effectiveness of this scheme.
\end{abstract}

%\renewcommand{\thefootnote}{\fnsymbol{footnote}}
%\footnotetext[1]{Send correspondence to Min Chen, E-mail: minchen@sjtu.edu.cn}
%\footnotetext[2]{Send correspondence to Zheng-Ming Sheng, E-mail: zmsheng@sjtu.edu.cn}
\maketitle

The exploiting of the high acceleration gradient in a laser
wakefield accelerator \mbox{(LWFA)} leads to a possible way
towards compact and low cost high energy accelerators
~\cite{Tajima1979,Pukhov2002,LuPRST2007,NakajimaCOL2013}. To
compete with the radio-frequency accelerators, the \mbox{LWFA} requires studies pursuing high injection qualities. Injection
means placing  certain amount of electrons into the accelerating
and focusing phase with proper velocities that they can be
accelerated continuously. It can be either pre-accelerating some
electrons to the phase velocity of the wakefield such as the
colliding pulse injection scheme~\cite{Esarey97a, Faure06,
Rechatin09}, or slowing down the wake so that some electrons can
catch up with the accelerating phase such as the density ramping
or transition injection scheme~\cite{Geddes08, Gonsalves11,
Bulanov98, ZengJPP2012, Li13}. Recently a new injection scheme
called the ionization-induced injection was proposed
~\cite{MinChen06, Oz07}. This scheme is attractive  due to its
simple experimental setup and has been experimentally demonstrated
by several groups recently~\cite{Clayton10, McGuffey10, Pak10}.
Generally the ionization-induced injection scheme utilizes the
higher ionization threshold of the K-shell of a high-Z gas (such
as nitrogen, oxygen or argon) mixed with a low-Z gas (usually
hydrogen or helium) to control the initial phase of the electrons
ionized from the K-shell.

\begin{figure}
  \begin{overpic}[width=0.3\textwidth]{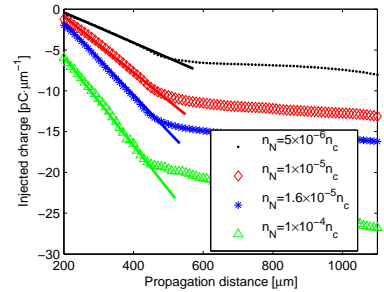}
  \end{overpic}
  \caption{\label{fig:charge_vs_distance_unmatched}(Color on line) Injected charge number vs laser propagation distance from 2D PIC simulations
  with an unmatched laser spot size. Only electrons in the first bucket are considered. The different symbols represent cases with different
  nitrogen atom density as shown in the legend. The solid lines are linear fittings for data with propagation distance from $200$ to $400~\rm \mu m$.
  The charge unit is $\rm pC \cdot \mu m^{-1}$ due to the 2D slab geometry. }
\end{figure}

Although the ionization-induced injection has the advantage of
relatively easy operation, it has large energy spread because the
ionization-induced injection usually occurs continuously until the
end of the mixed gas or beam loading effects
occur~\cite{Tzoufras08, Chen2012}. Efforts have been devoted to
reduce the mixed gas length. In 2011, Liu \textit{et al.} reduced
the mixed gas length to $1\rm mm$ by the two-stage accelerating
configuration~\cite{Liu2011, Pollock11}. In this experiment the
first stage, called injector which filled with mixed gas, is about
$1\rm mm$ long. The following stage, called accelerator which
filled with pure He, provides longer distance acceleration without
injection. By this injector-accelerator scheme, a final absolute
energy spread about 50MeV is observed. To further reduce the
absolute energy spread one needs to reduce the mixed gas length
further. However, the up-to-date minimum size of the mixed gas jet
is about millimeter scale, which is still longer than
satisfactory. Xia \textit{et al.} demonstrated the
quasi-monoenergetic beam generation by controlling the amount of
charge from ionization-induced injections by using the laser power
close to the ionization threshold of the inner shell of
oxygen~\cite{XiaPOP2011}. In their work, the injected electron
energy spread is optimized to about $50 \rm MeV$ by adjusting the
laser peak intensity. The process is both self-focusing and
pump-depletion controlled and thus the optimization is not
satisfactory. Other efforts such as using separated lase pulses to
cut the gas profile are also devoted~\cite{Hsieh2006}. Although
these experiments can improve the final beam quality, usually
complicated experimental configurations are required. All optical
cutting in ionization based electron injection in LWFA and low
energy spread beam production has not been reported.

In this letter, we propose a simple optical method to
cut the injection length down to a few hundred micron meters
with single stage only, which is much shorter than the
mechanical limits obtained so far. This method utilizes the
self-focusing process~\cite{Sun87, Mori97} of an initially
unmatched laser pulse which can automatically truncate the
injection process due to bubble deformation before the mixed gas ends.
Other than in Ref~\cite{XiaPOP2011} which use self-focusing to start the ionization-induced injections, the self-focusing of
the driver pulse in our work leads to a strong wakefield evolution which later
breaks the ionization-induced injection condition and suppresses
the injection process. The resulted injection length is shown to be determined by the period of the laser spot shrinks to the minimum size, which is
usually a few hundred microns. This provides a simple way for high quality (low
energy spread) ionization-induced injections.

In our simulations, the gas target is composed of nitrogen and
helium mixed gas.  The mixed gas is initially uniformly
distributed except the $200\rm \mu m$ up-ramp from the vacuum to the gas. The simulations are performed in
OSIRIS~2.0 framework with the moving window scheme so that the gas target can be half-infinite~\cite{OSIRIS}. The simulation box size
is $50 \times 100~\rm{\mu m}^2$, the cell size is $0.015625 \times
0.25~\rm{\mu m}^2$ and the time step interval is $0.05~\rm fs$.
The helium density is kept to be $1.6\times 10 ^{-3} n_{\rm c}$ in all simulations,
where $n_{\rm c} = 1.745 \times 10^{21} \rm cm^{-3}$ is the critical density for the $800 \rm nm$ laser. The nitrogen atom
density ($n_{\rm N}$) is relative small and varies from
$5\times10^{-6} n_{\rm c}$ to $1\times10^{-4} n_{\rm c}$. Due
to the small concentration of nitrogen, the beam loading effects of
the injected electrons are negligible, which will be explained in
details in the following. The laser pulse is S-polarized in
two-dimensional (2D) slab geometry simulations with wavelength of $0.8\rm \mu m$ and pulse duration
of $L_{\rm FWHM}=33~\rm fs$. The laser amplitude and focal waist
parameters vary either at the so-called matched or unmatched
condition for self-guiding~\cite{LuPRST2007}.

To study the injection process,  we first study the charge
injection rate along the laser propagation distance in 2D-PIC
simulations. The laser beam has a normalized vector potential of
$a_0 = eA/mc^2=2.0$ and the waist of $k_{\rm p}W_0 = 7.594$.
Fig.~\ref{fig:charge_vs_distance_unmatched} shows the injected
beam charges vs laser propagation distance from simulations with
different nitrogen concentration. One may notice in all the cases
the injected beam charge saturates almost over the same laser
propagation distance, which is around $400\sim450~\rm{\mu m}$.
This phenomenon does not agree with the common understanding of
the ionization-induced injection process. Usually
ionization-induced injection will be continuous if the mixed gas
length is long enough, or it will stop due to beam loading
effects. According to the equation of maximum affordable number of electrons in the bubble regime $N \simeq 2.5 \times 10^9 \lambda[{\rm \mu m}] / 0.8 \cdot \sqrt{P[{\rm TW}]/100}$~\cite{LuPRST2007}, the beam loading should occur when the charge approaches $249 \rm pC$ (or $36 \rm pC / \mu m$ in 2D slab geometry if we assume the width of the beam is about $7 \rm \mu m$, which is a common beam width). In our case, the charge is proportional to the
nitrogen concentration. This proves that the injected charges (at least
for the low nitrogen concentration cases) are lower than that can trigger the beam loading effect. So the only reason for this
kind of injection truncation is due to the shortening of the
effective injection length by other mechanisms.
\begin{figure}
\begin{tabular}{lc}
  \begin{overpic}[width=0.23\textwidth, trim=30 0 10 0]{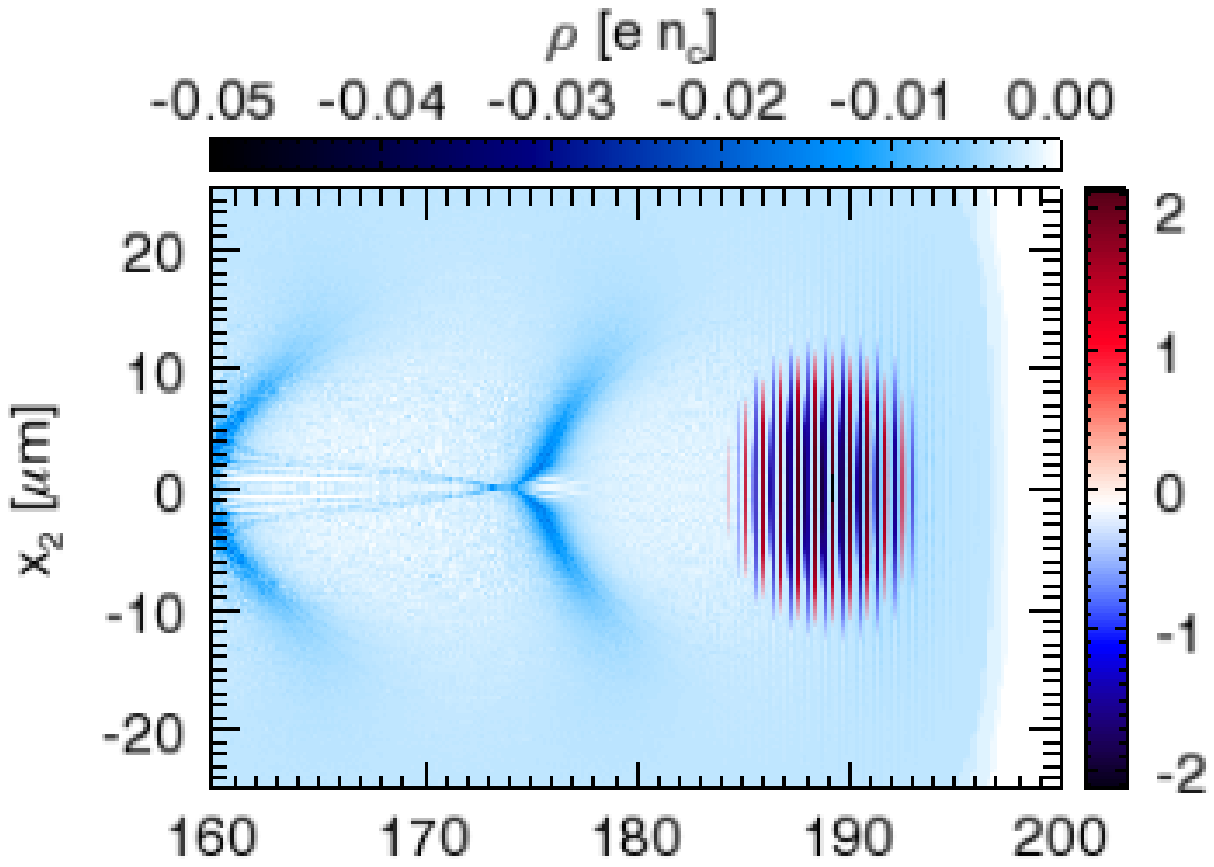}
    \put(20,17){(a)}
  \end{overpic}
  \begin{overpic}[width=0.23\textwidth, trim=30 0 10 0]{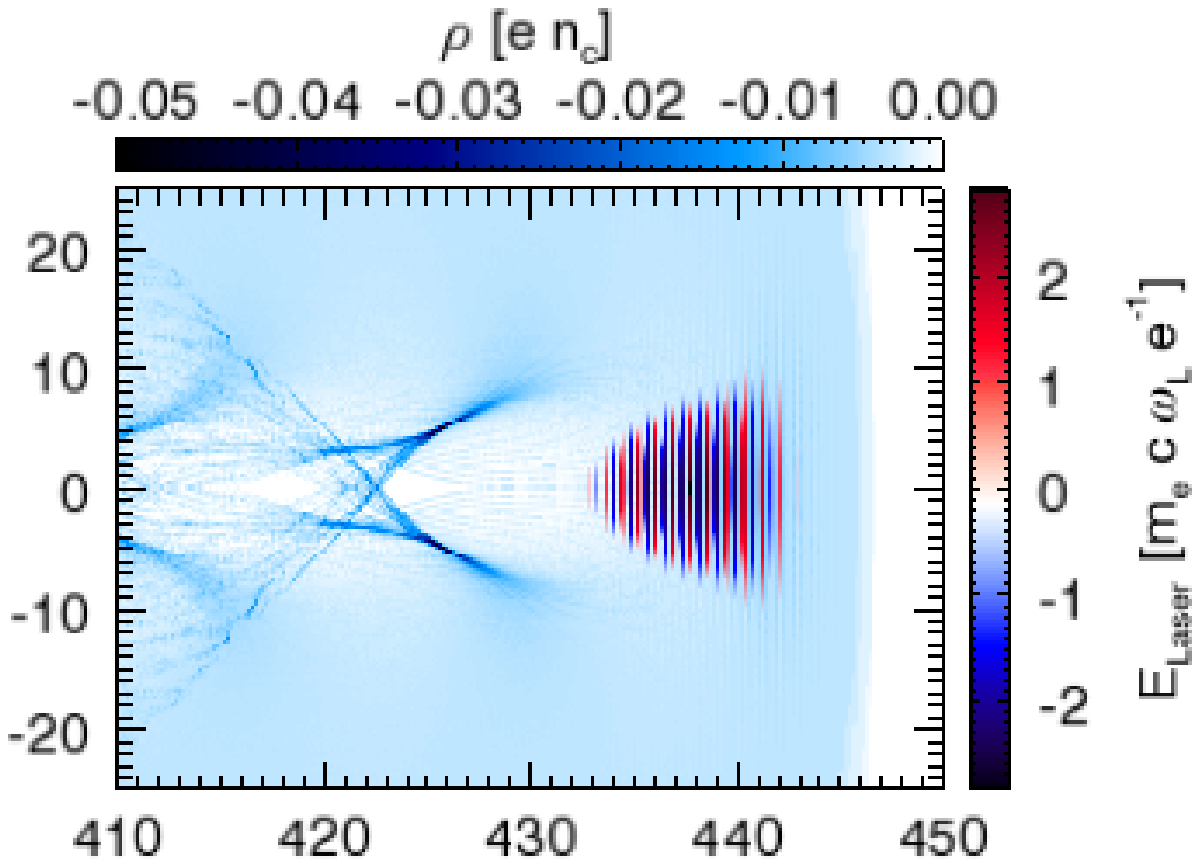}
    \put(16,17){(b)}
  \end{overpic} \\
  \begin{overpic}[width=0.23\textwidth, trim=0 0 0 0]{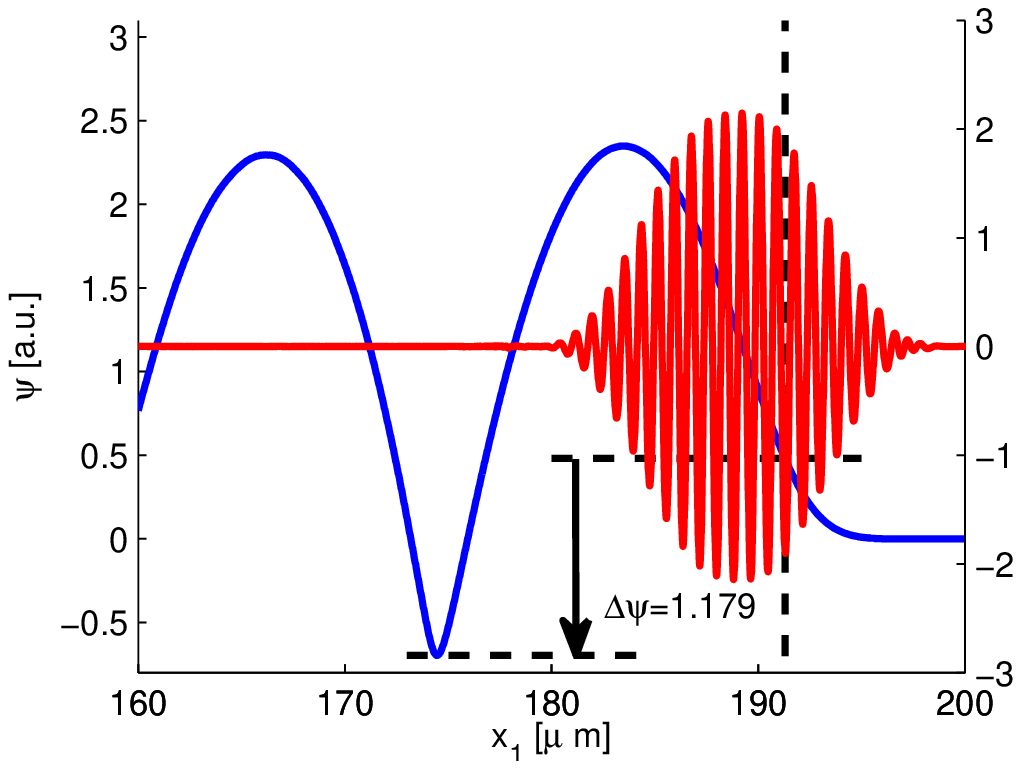}
    \put(19,62){(c)}
  \end{overpic}
  \begin{overpic}[width=0.23\textwidth, trim=0 0 0 0]{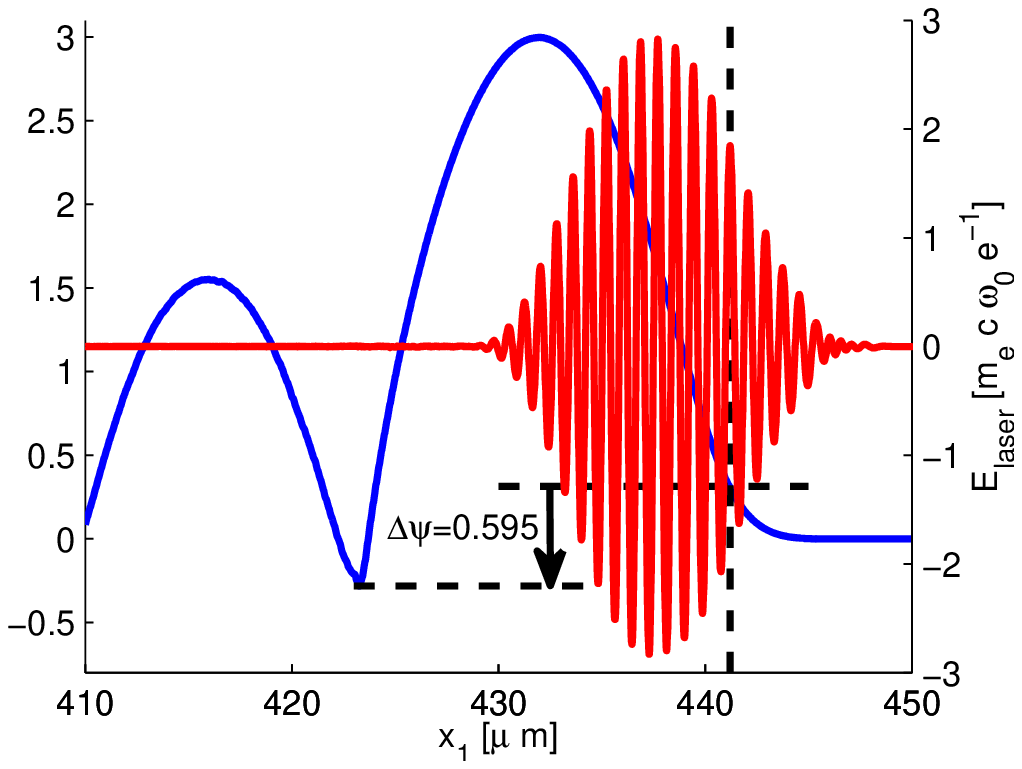}
    \put(17,62){(d)}
  \end{overpic}
\end{tabular}
  \caption{(Color on line)\label{fig:laser_psi}
    (a, b) Electron density and laser electric field distribution at the laser propagation distance of $200$ and $450~\rm{\mu m}$, respectively.
    (c, d) Laser (red lines) and pseudo-potential (blue lines) line-outs at these two different time steps. Densities are normalized by
    $n_{\rm c} = 1.745 \times 10^{21}~\rm cm^{-3}$ and electric fields are
    normalized by $E_0 = m_e c \omega_{\rm L} e^{-1} = 4 \times 10^{12}~ \rm{V/m}$. The dashed lines show the ionization starting point of nitrogen
    inner shell and the fall to the bottom of the potential well ($\Delta \psi = 1.179$ and $0.595$, respectively).
  }
\end{figure}

In order to see what induces the injection and cutoff processes,
we plot the wake,  laser field and the pseudo-potential of the
wakefield at two different time steps in Fig.~\ref{fig:laser_psi}.
From the static model of plasma based
accelerators~\cite{LuPRL2006, Pak10, Chen2012}, one knows that the
ionization-induced injection occurs when the pseudo-potential
difference between the electron ionization position and the end of
the wake bucket satisfies $\Delta \psi \geq 1$.
Figure~\ref{fig:laser_psi}(a, c) show that at the first selected
time step the laser pulse still holds its gaussian shape and $\Delta
\psi$ of the excited wake is large enough for the
ionization-induced injection to occur. However at the second
selected position ($450~\rm{\mu m}$) the
laser is strongly deformed and self-focused, and the profile is transformed to a
bell-like shape (front steepened), meanwhile the laser field amplitude is
about $45\%$ higher than initial. This larger amplitude laser pulse excites a wake with a larger amplitude
(see Fig.~\ref{fig:laser_psi}(d)). The minimum potential increases which
makes $\Delta \psi$ smaller and the ionization-induced injection
condition ($\Delta \psi \ge 1$) is no longer satisfied. This is the reason of the injection truncation. Since the nitrogen concentration in all
of our simulations is low, the laser pulse evolution is mainly
affected or determined by the background plasmas. The pulse
evolution in all of the simulations are almost the same. So the
truncation positions of the ionization-induced injections are
almost the same regardless of the change of $n_{\rm N}$.

The above analysis shows that it is the self-evolution of the
pulse terminates the ionization-induced injection process.  As we
know the pulse evolution in the plasma depends on its initial
power and focus state. A matched laser pulse can have a very
stable transverse profile during its propagation. To check our
analysis we make a serials of simulations which keep the pulse
power in 2D geometry but varying the pulse's initial transverse
focusing size, which affects the pulse's self-evolution.

\begin{figure}
\begin{tabular}{cc}
  \begin{overpic}[width=0.24\textwidth]{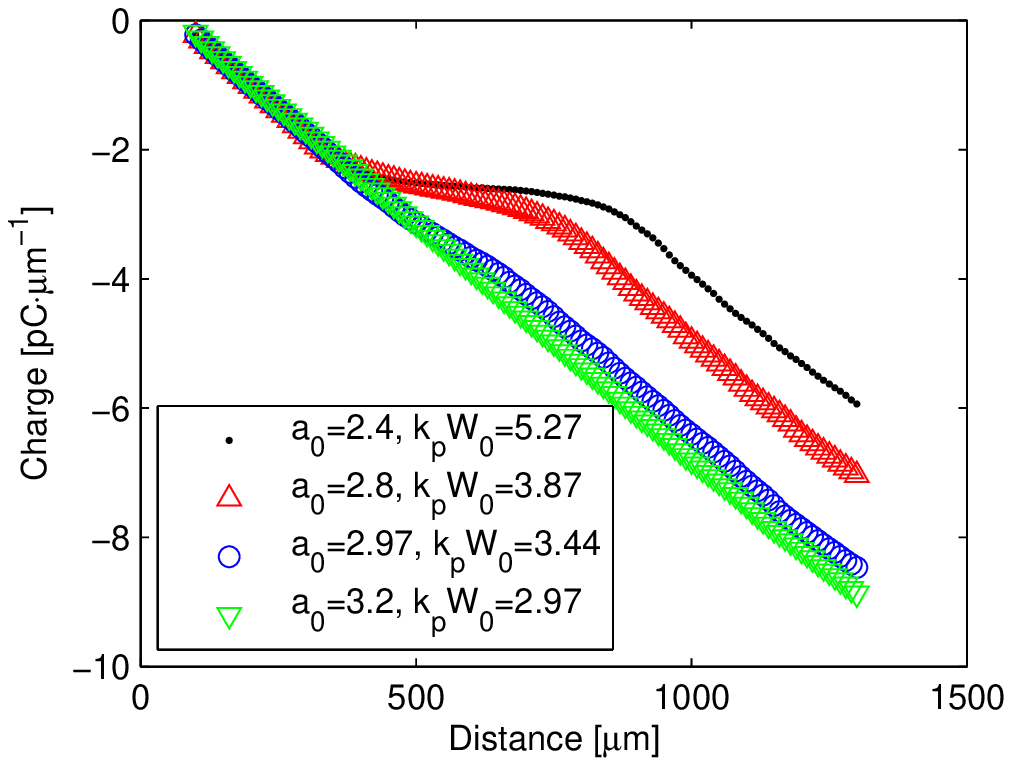}
    \put(75,60){\bf \textcolor{black}{(a)}}
  \end{overpic} &
  \begin{overpic}[width=0.24\textwidth]{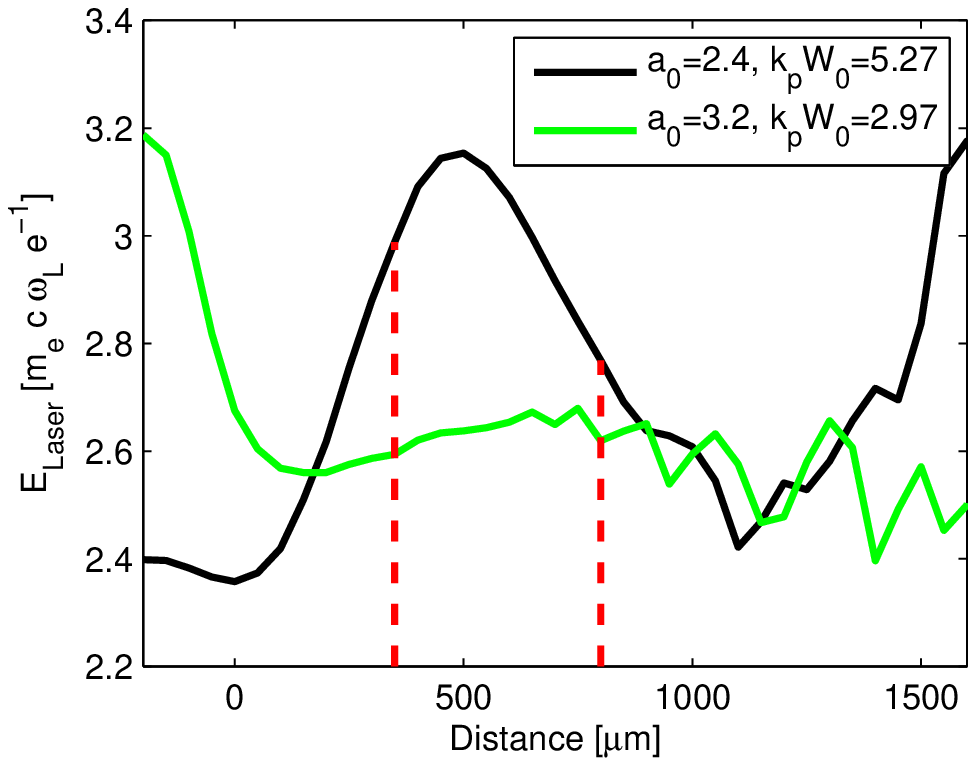}
    \put(20,60){\bf \textcolor{black}{(b)}}
    \put(45,13){\begin{sideways}\textcolor{red}{\tiny \fontfamily{phv}\selectfont Injection suppressed}\end{sideways}}
  \end{overpic}
\end{tabular}
  \caption{(Color on line)\label{fig:charge_vs_distance330_333}
    (a) Injected charge number vs. laser propagation distance from 2D PIC simulations with different normalized laser
    vector potential $a_0$ and normalized waist $k_{\rm p}W_0$. The laser energies are kept the same in 2D slab geometry by keeping $a_0^2k_{\rm p}W_0$  a constant.
    (b) Maximum axial laser field evolution of two cases: black line for $a_0=2.4$, $k_{\rm p}W_0=5.27$ and green line for $a_0=3.2$, $k_{\rm p}W_0=2.97$.
    The red dashed lines indicate the region where the injection is suppressed for the black line case.
  }
\end{figure}

The matched spot size in relativistic condition (estimated with
large $a_0$),  described in Ref.~\cite{LuPRST2007}, is $k_{\rm p}
W_0 = 2 \sqrt{a_0}$, where $k_{\rm p}$ is the wave number of a
plasma wave. Earlier 3D simulations show a laser beam satisfying
this matching condition can keep its shape until pump depletion
develops. In the following, we compare the ionization induced
injection rate between different laser parameters, i.~e.~the
normalized vector potential $a_0$ and waist $k_{\rm p}W_0$. The
ratio $n_{\rm N}/n_{\rm H_e}$ is fixed to be $0.1\%$ so that the beam
loading effect is negligible. In order to keep the laser power in
2D slab geometry, we keep $a_0^2W_0$ to be a constant in different
simulation cases, which could be relevant for an asymmetric spot
size in 3D.

\begin{figure}
\begin{tabular}{lc}
  \begin{overpic}[width=0.24\textwidth, trim=0 0 20 0]{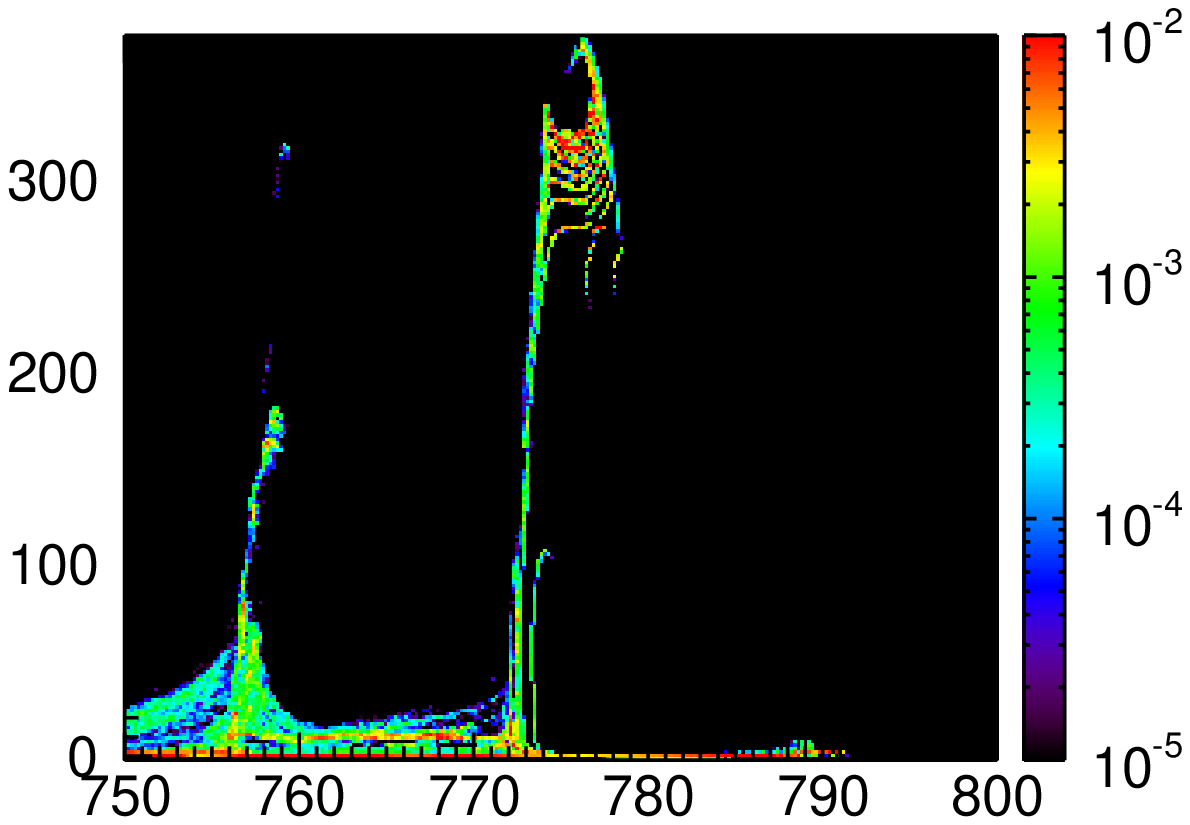}
    \put(23,60){\bf \textcolor{white}{(a)}}
    \put(43,0){\scriptsize \fontfamily{phv}\selectfont {$x_1 \left[\rm \mu m\right]$}}
    \put(3,27){\begin{sideways}\scriptsize \fontfamily{phv}\selectfont {$p_1 \left[\rm m_ec\right]$}\end{sideways}}
  \end{overpic}&
  \begin{overpic}[width=0.24\textwidth, trim=40 0 -20 0]{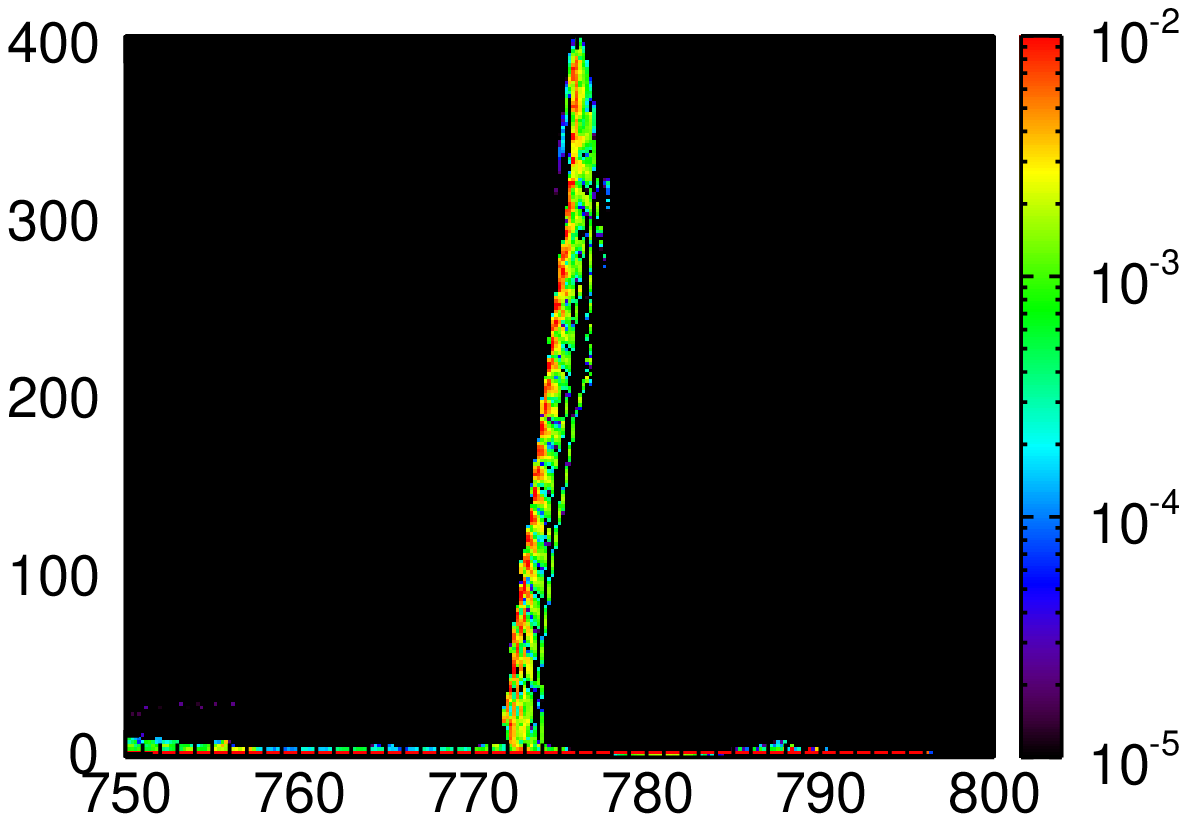}
    \put(13,60){\bf \textcolor{white}{(b)}}
    \put(33,0){\scriptsize \fontfamily{phv}\selectfont {$x_1 \left[\rm \mu m\right]$}}
    \put(92,0){\begin{sideways}\scriptsize \fontfamily{phv}\selectfont {Phase space density $\left[\rm a.u.\right]$}\end{sideways}}
  \end{overpic}\\

  \begin{overpic}[width=0.215\textwidth, trim=0 0 19 0]{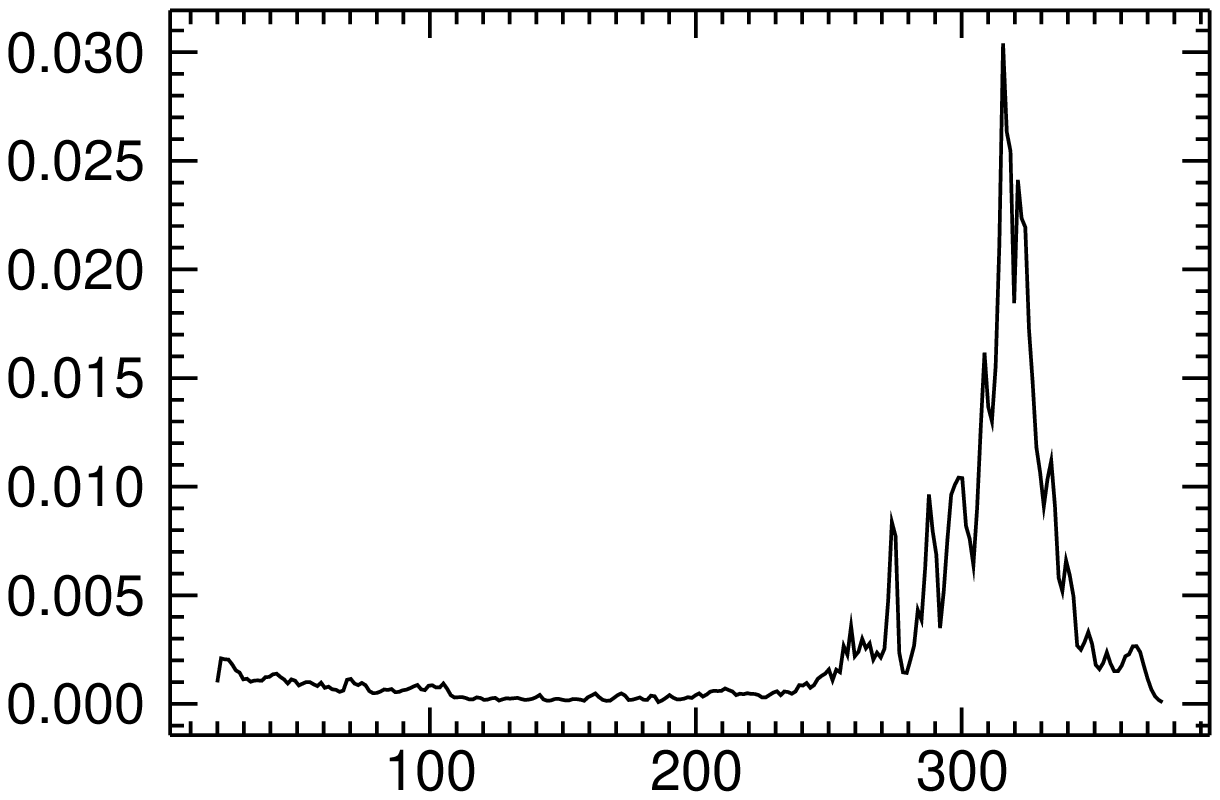}
    \put(26,58){\bf \textcolor{black}{(c)}}
    \put(3,22){\begin{sideways}\scriptsize \fontfamily{phv}\selectfont {$\rm Count \left[\rm a.u.\right]$}\end{sideways}}
    \put(57,0){\scriptsize \fontfamily{phv}\selectfont {$\gamma-1$}}
  \end{overpic}&
  \begin{overpic}[width=0.215\textwidth, trim=59 0 -40 0]{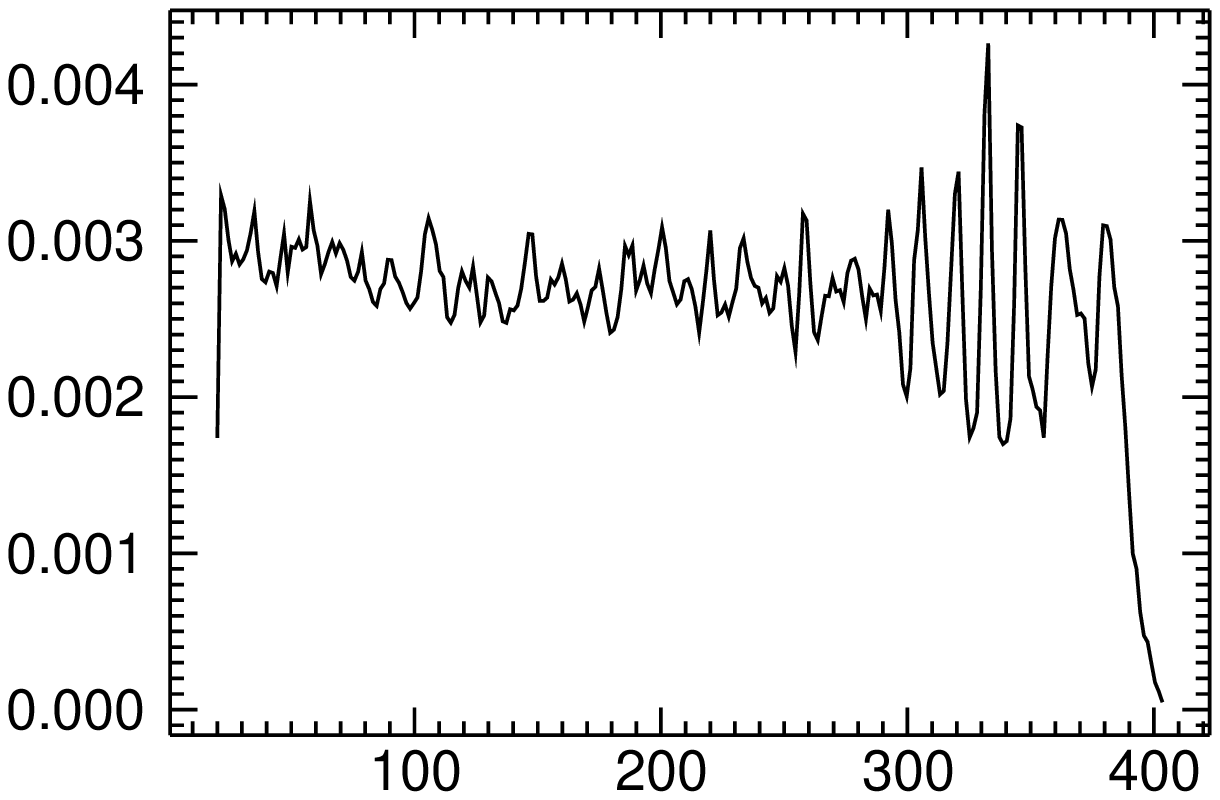}
    \put(12,58){\bf \textcolor{black}{(d)}}
    \put(45,0){\scriptsize \fontfamily{phv}\selectfont {$\gamma-1$}}
  \end{overpic}\\

\end{tabular}
  \caption{(Color on line)\label{fig:phase_spec_330_333}
    The phase space (a, b) and the energy spectra (c, d) of the injected electrons for the cases (a, c) $a_0=2.4$, $k_{\rm p}W_0=5.27$ and
    (b, d) $a_0=3.2$, $k_{\rm p}W_0=2.96$, i.e. the black dots and the green triangle in Fig.~\ref{fig:charge_vs_distance330_333}~(a), respectively.
    The beam qualities are (a, c) $7.1~\rm pC$ in charge, $1.05~\rm mm \cdot mrad$ in emittance and $7.9\%$ energy spread in FWHM with central energy of $160~\rm MeV$,
    (b, d) $15.9~\rm pC$ in charge, $1.1~\rm mm \cdot mrad$ in emittance and $100\%$ energy spread with maximum energy of $200~\rm MeV$.
  }
\end{figure}

Figure~\ref{fig:charge_vs_distance330_333}(a) shows the
ionization-induced injection rates with different laser
parameters.  The case $a_0=2.4$, $k_{\rm p}W_0=5.27$ shows a clear
injection suppression start from about $400~\rm\mu m$. The power
to self-focusing critical power ratio is $P/P_c =\alpha ( k_{\rm
p}W_0a_0)^2/32 = 7.07$ in this case ($\alpha = \sqrt{2}$ for 2D
slab geometry)~\cite{TzengPRL1998}. The case $a_0=2.97$, $k_{\rm
p}W_0=3.44$ is the best matched case from the theory, but the case
$a_0=3.2$, $k_{\rm p}W_0=2.96$ shows a better linear injection
behavior. This discrepancy is because  the above matching
condition is an estimation with large $a_0$.
Figure~\ref{fig:charge_vs_distance330_333}(b) shows the evolution
of on-axis laser electric field. As one can see, in the most
unmatched case with $a_0=2.4$ the ionization-induced injection
happens periodically. This is due to the pulse's periodically
focusing and defocusing and the consequent bubble evolution. The
two red dashed vertical lines labels the region where
ionization-induced injection is suppressed. In this simulation two
bunches of ionization-induce electron injections can be seen in
the same wake bucket. The final spectrum shows one mono-energetic
peak with a low energy background, which has been observed in
serval former experiments. However, in most of the reports, this
phenomenon is explained by regarding the electrons to be in
different buckets.

\begin{figure}
\begin{tabular}{cc}
  \begin{overpic}[width=0.24\textwidth]{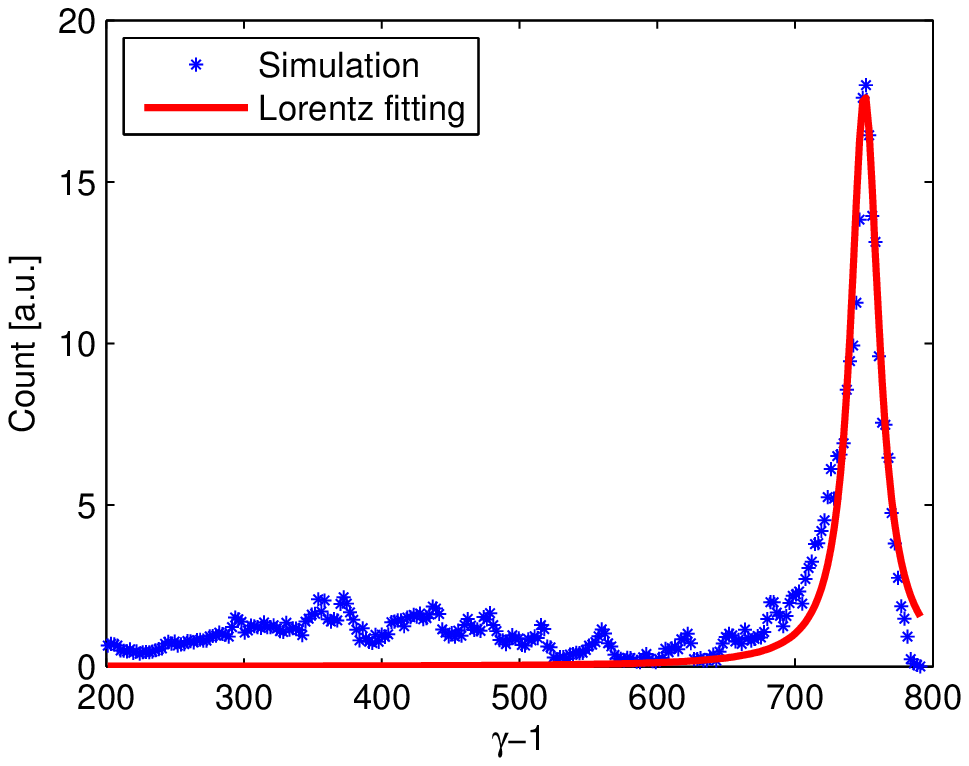}
    \put(15,50){\bf \textcolor{black}{(a)}}
  \end{overpic}&
  \begin{overpic}[width=0.24\textwidth]{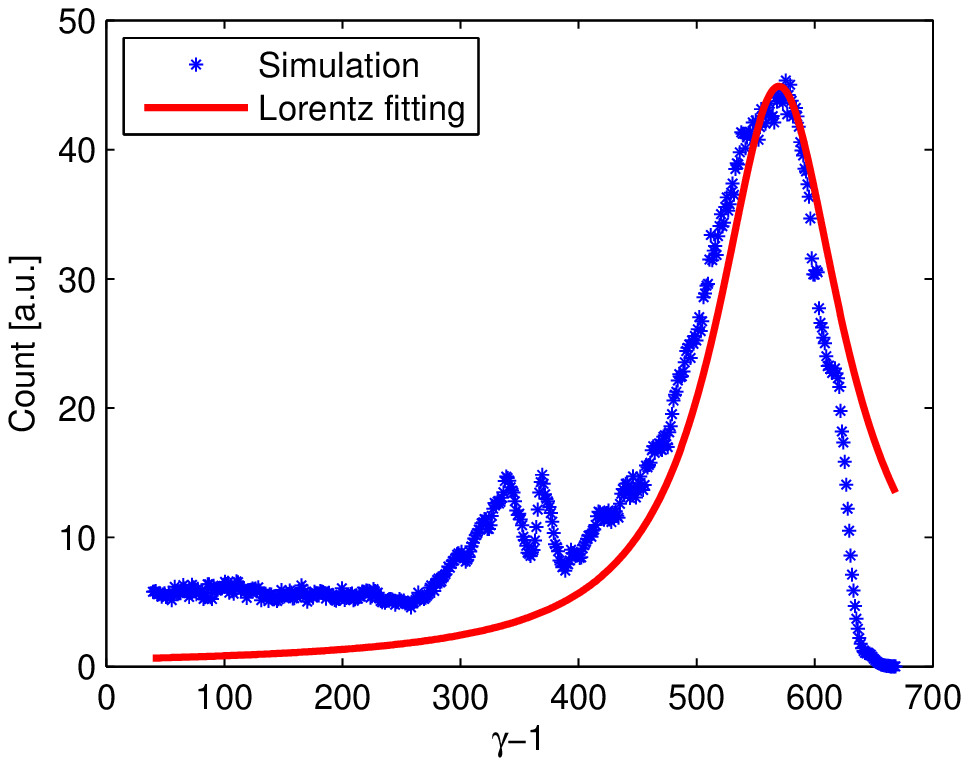}
    \put(15,50){\bf \textcolor{black}{(b)}}
  \end{overpic}\\
  \begin{overpic}[width=0.24\textwidth, trim=-8 0 8 0]{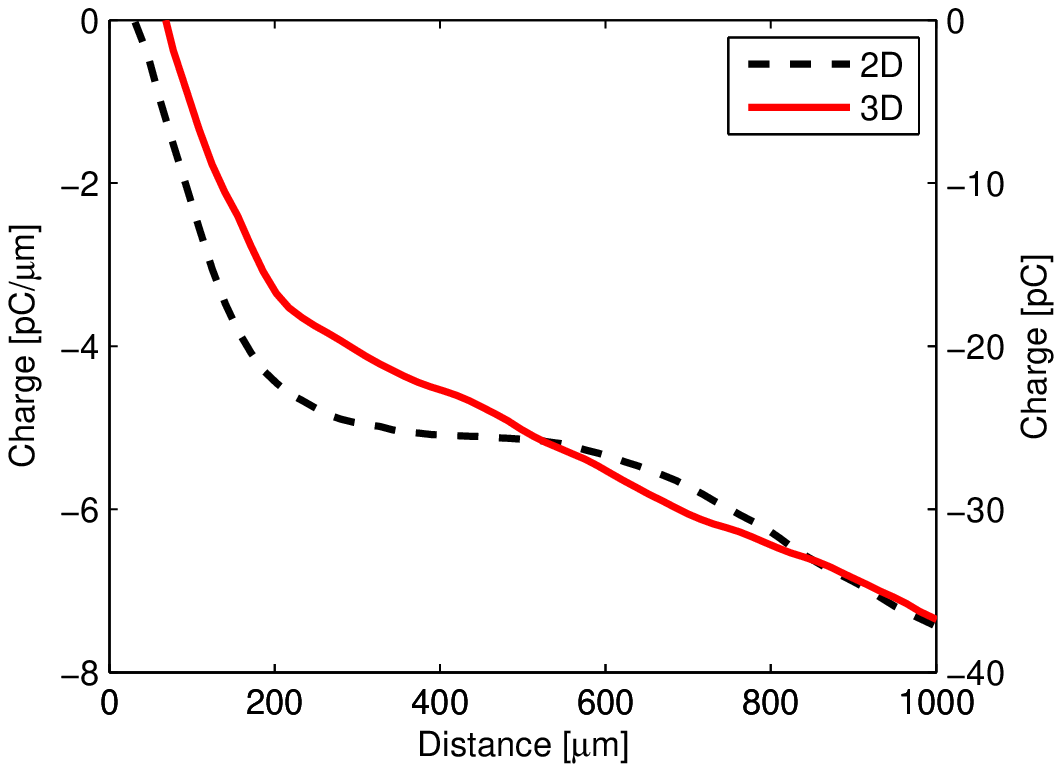}
    \put(15,13){\bf \textcolor{black}{(c)}}
  \end{overpic}&
  \begin{overpic}[width=0.25\textwidth]{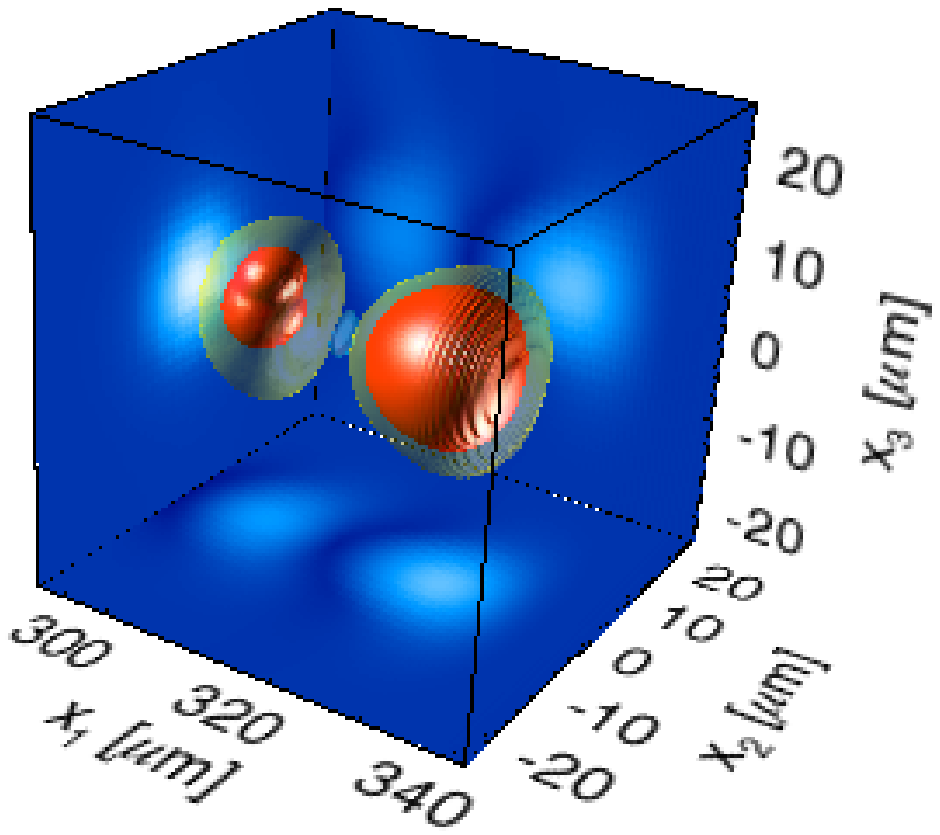}
    \put(15,63){\bf \textcolor{black}{(d)}}
  \end{overpic}
\end{tabular}
  \caption{(Color on line)\label{fig:optimized}
    The optimized 2D simulation and its comparison with the 3D simulation.
    (a) The optimized 2D simulation result with laser parameters $a_0=2.9$ and $k_{\rm p}W_0=7.594$.
    (b) The 3D simulation beam energy spectrum with the same parameters as in the optimized 2D case. The spectrum is taken at propagation distance of $760 \rm \mu m$.
    (c) The injected beam charge vs.\ propagation distance for the 2D simulation shown in (a) and 3D simulation shown in (b). When calculating the propagation distance, the density transition between vacuum and the gas is excluded. The charge zero point is not the distance zero point because it takes a while for the electrons to reach the injection criteria.
    (d) The pseudo-potential $\psi$ isosurface and its projections for the 3D wake, which evolves to a nut-like shape at propagation distance of a few hundred microns. The 3D simulation has the same simulation box size as 2D simulations, but the displayed region is reduced.
  }
\end{figure}

We compare the phase space and energy spectrum for these two cases
in Fig.~\ref{fig:phase_spec_330_333}. With the unmatched laser
pulse, the injected charge is reduced by a half but the energy
spread is greatly improved, and mono-energetic electron beam due
to ionization injection can be seen (Fig.~\ref{fig:phase_spec_330_333}(a, c)). Simulations prove that
slightly increasing the $n_{\rm N}$ can recover the final injected
charge but keep the low energy spread in unmatched cases. In the
range of mixing ratio $<1\%$,  the injected charge increases
almost linearly with $n_{\rm N}$. In the matched case, due to the
continuous injection, accelerated electron distribution in phase
space appears continuous. This results in an almost flat spectrum
in the final energy distribution (see
Fig.~\ref{fig:phase_spec_330_333}(b, d)).

By estimating the self-focusing length, one can optimize the
injection for better beam quality.  To do this, we consider the
equation for laser profile evolution with $d^2 R/dz^2 = Z_{\rm
R}^{-2} R^{-3} (1-P/P_{\rm c})$, where $R=W/W_0$ is the normalized
spot size and $Z_{\rm R}=kW_0^2/2$ is the vacuum Rayleigh length
~\cite{TzengPRL1998, EsareyRMP2009}. One can obtain the analytic
solution $R^2=1+(1-P/P_{\rm c})z^2/Z_{\rm R}^2$, which indicates
that the laser spot size shrinks to 0 at $z=z_{\rm cut}\equiv
Z_{\rm R}(P/P_{\rm c}-1)^{-1/2}$, where $P/P_{\rm c} =\alpha ( k_{\rm
p}W_0a_0)^2/32$. We can estimate the injection length to be
$z_{\rm cut}$. The former laser parameters ($a_0=2$ and $k_{\rm
p}W_0=7.594$) gives $z_{\rm cut} = 373 \rm \mu m$, which is
consistent with previous simulation
(Fig.~\ref{fig:charge_vs_distance_unmatched}). Next, we keep the
laser power to be the same ($P=39\rm TW$, thus $a_0k_{\rm p}W_0 =
15.2$) and change the initial $a_0$. From a series of simulations
we found that the ionization-induced injections can be
sufficiently reduced if $a_0>2.9$. Thus if we choose the condition
$a_0=2.9$, the minimum injection length is obtained to be $178 \rm
\mu m$. The PIC simulations also confirm this estimation (the
injection length is reduced to $187\rm \mu m$ with this set of
parameters). At this condition, the ionization injection leads to
the generation of an electron beam with $14.58 \rm pC$ in charge,
$383\rm MeV$ in the central energy with energy spread of $\Delta
E_{\rm FWHM}/E = 3.33\%$ (shown in Fig.~\ref{fig:optimized}~(a)), and normalized emittance of $3.12 \rm
mm\cdot mrad$ at the propagation distance $1.4\rm mm$. In this
optimized case, the initial laser parameters are $a_0 = 2.9$ and
$W_0=11.69 \rm \mu m$, and the gas is $n_{\rm H_e}=2.8\times
10^{18} {\rm cm}^{-3}$ and $n_{\rm N}=8.5\times 10^{15} {\rm
cm}^{-3}$ mixed. Compared with a very recent published work on
two-pulse ionization injection~\cite{BourgeoisPRL2013}, our work
has the advantages on simpler configuration using one pulse only
and larger charge with similar beam emittance.

In a real case, the pulse evolution is actually a
three-dimensional (3D) process, therefore 3D simulation is
necessary to check if self-truncated ionization injection really
exists. The results from the 3D simulation with the same laser
parameters as in the 2D optimized case are shown in
Fig.~\ref{fig:optimized}~(b-d). One may notice that in 3D the beam
energy spread is larger than in 2D,  and the ionization-induced
injection is slowed down at $200\rm \mu m$ instead of totally
suppressed (Fig.~\ref{fig:optimized}~(c)). This is because the 3D symmetric laser pulse has
different evolution compared with that in 2D-slab geometry.
Nevertheless, the quasi-monoenergetic character still confirms the
validity of this injection truncation phenomenon. More simulations
towards 3D optimization is our future goal.

In summary we have demonstrated a simple method to shorten the
effective injection length in the ionization based scheme.  The
self-evolution of an initially unmatched laser pulse can break the
ionization-induced injection condition and make the injection
length much shorter than that for matched laser beam, which makes
the final accelerated electron beam  mono-energetic. Different
from the former researches of the self-injection due to the bubble
evolution~\cite{Kalmykov2009,Corde2013}, the injection here is
ionization-induced. It deserves to point out that since the
shortening of the injection is based on pulse evolution, long
distance acceleration for achieving high energy acceleration
requires matched spot size at further stages. Nevertheless, our
study is helpful for the initial injection process, which is vital
for final beam qualities.

% Put your acknowledgments here.
The authors would like to acknowledge the OSIRIS Consortium,
consisting of UCLA and IST (Lisbon, Portugal) for the use of
OSIRIS and the visXD framework. MZ appreciates the useful
discussions with Frank Tsung, Weiming An, Asher Davidson, Feiyu Li
and Yue Liu. This work is supported by the National Basic Research
Program of China (Grant No. 2013CBA01504) and the National Science
Foundation of China (Grants No. 11121504, 11129503, 11075105 and
11205101). Simulations were performed on the Hoffman2 at UCLA,
Hopper at NERSC, and Supercomputer $\Pi$ at Shanghai Jiao Tong
University.

\bibliography{May2013}
\end{document}